\documentclass[journal]{IEEEtran}
\usepackage{amsmath,amssymb,graphicx,amsfonts,graphics,bbm,color}
\usepackage [english]{babel}
\newcommand{\ket}[1]{\ensuremath{\left|#1\right\rangle}}
\newcommand{\bra}[1]{\ensuremath{\left\langle#1\right|}}
\def\QBER{\hbox{QBER}}
\def\LLR{\hbox{LLR}}

\newcommand{\STErev}[1]{{#1}}
\begin{document}
\title{Soft-metric based channel decoding for photon counting receivers }
\author{
Marina~Mondin,
\thanks{M.~Mondin is with the Dipartimento di Elettronica e
Telecomunicazioni, Politecnico di Torino, 10129 Torino, Italy, e-mail: marina.mondin@polito.it} 
Fred~Daneshgaran,
\thanks{F.~Daneshgaran is with the Electrical and Computer Eng. Dept., California State Univ., Los Angeles, CA, USA, e-mail: fdanesh@calstatela.edu}
Inam~Bari,
\thanks{
I. Bari is with the National University of Computer and Emerging Sciences
(FAST-NU), Peshawar, Pakistan e-mail: inam.bari@nu.edu.pk}
Maria~Teresa~Delgado,
\thanks{M.~T.~Delgado is with the Dipartimento di Elettronica e
Telecomunicazioni, Politecnico di Torino, 10129 Torino, Italy, e-mail: maria.delgadoalizo@polito.it}\\
Stefano~Olivares,
\thanks{S.~Olivares is with the Dipartimento di Fisica, Universit\`a degli Studi di Milano, I-20133 Milano, Italy, e-mail: stafano.olivares@fisica.unimi.it}
and Matteo~G.~A.~Paris
\thanks{M.~G.~A.~Paris is with the Dipartimento di Fisica, Universit\`a degli Studi di Milano, I-20133 Milano, Italy, e-mail: matteo.paris@fisica.unimi.it}}
\markboth{Journal of \LaTeX\ Class Files}%
{Mondin \MakeLowercase{\textit{et al.}}: Soft-metric based channel decoding for photon counting receivers}
\maketitle 
%%%
\begin{abstract}
We address photon-number-assisted, polarization-based, binary
communication systems equipped with photon counting receivers.  In these
channels information is encoded in the value of polarization phase-shift
but the carrier has and additional degree of freedom, i.e.  its photon
distribution, which may be exploited to implement binary input-multiple
output (BIMO) channels also in the presence of a \STErev{phase-diffusion
noise affecting the polarization}.  Here we analyze the performances of
these channels, which approach capacity by means of iteratively decoded
error correcting codes.  In this paper we use soft-metric-based low
density parity check (LDPC) codes for this purpose.  In order to take
full advantage of all the information available at the output of a
photon counting receiver, soft information is generated in the form of
log-likelihood ratios, leading to improved frame error rate (FER) and
bit error rate (BER) compared to binary symmetric channels (BSC). We
evaluate the classical capacity of the considered BIMO channel and show
the potential gains that may be provided by photon counting detectors in
realistic implementations.  
\end{abstract}


\IEEEpeerreviewmaketitle

%%%
\section{Introduction}
%%%
\IEEEPARstart{I}n binary 
optical communication, the logical information is encoded onto two
different states of the radiation field. After the propagation, the
receiver should perform a measurement, aimed at discriminating the two
signals. Currently, most of the long-distance amplification-free optical
classical communication schemes employ relatively weak laser sources
leading to small mean photon count values at the receiver. The same is
true for quantum-enhanced secure cryptographic protocols.  In fact,
laser radiation, which is described by coherent states, preserves its
\STErev{Poissonian photon-number statistics and polarization} also in
the presence of losses. On the other hand, operating in the regime of
low number of detected photons gives rise to the problem of
discriminating the signals by quantum-limited measurements
\cite{lab1,lab2,lab2a}. Indeed, the binary discrimination problem for coherent
states has been thoroughly investigated, both for its fundamental
interest and for practical purposes
\cite{lab3,lab4,lab5,lab6,lab7,lab8}.  It should be mentioned however
that in order to exploit the phase properties of coherent states, one
should implement phase sensitive receivers \cite{lab9,lab10} with nearly
optimal performances also in the presence of dissipation and noise
\cite{lab6,lab11}. This is a challenging task, since it is generally
difficult, and sometimes impossible, to have a suitable and reliable
phase reference in order to implement this kind of receiver
\STErev{\cite{bina:ken}}.
\par
The simplest choice for a detection scheme involving radiation is given
by detectors which simply reveal the presence or the absence of
radiation (on/off detectors) with acceptable dead-time values and dark
count rates. A natural evolution of such schemes would be to employ
photon counting receivers.  Indeed, development of photon counters has
been extensively pursued in the last decades, as well as of methods to
extract the photon distribution by other schemes
\cite{rossi:04,bondani:09,allevi:09,allevi:10,kala:11}.  Given that one
could use photon counting detectors for weak-energy optical
communications, a question arises on whether and how such detectors may
be employed to improve the system performance. A possible way to answer
this question is to determine the capacity of the corresponding optical
channels, and the achievable residual Bit Error Rate (BER) and Frame
Error Rate (FER) of practical communication schemes over these channels.
A photon counting detector is clearly able to extract more information
than a simple on/off detector.  The practical consequence is that a
photon counting detector allows one to generate a meaningful
log-likelihood (i.e. a soft-metric), as opposed to a hard-metric allowed
by a hard- (or on/off) detector.  Furthermore, soft-metrics lead to
improved performances when exploited by powerful
iteratively decoded forward error correcting codes.
\par
Recently, a \STErev{simple} polarization-based communication scheme
involving weak coherent optical signals and low-complexity photon
counting receivers has been presented \cite{lab1}, and its performances
have been analyzed based on an  equivalent Binary Symmetric Channel
(BSC) model of the overall scheme. In this paper, we extend the scheme
of \cite{lab1} and model the effect of the photon distribution of the
coherent signals as a time varying Binary
Input-Multiple Output (BIMO) channel.  In particular, we employ
soft-metric based Low Density Parity Check (LDPC) codes for transmission
over the BIMO channel to approach capacity using iteratively decoded
error correcting codes and investigate the potential improvements that
may be obtained in terms of classical capacity and residual BER using
photon counting receivers \cite{lab18}. \STErev{It is worth noting that
recently photon-counting detectors have been proposed to enhance the
discrimination of weak optical signal in the case of $M$-ary coherent
state discrimination \cite{refcite1,refcite2}: in these cases, however,
a suitable feedback scheme or the use of squeezing are required.} 
\par 
The receiver introduced in \cite{lab1} is based on an optical setup for
one-parameter qubit gate optimal estimation \cite{lab2,lab12,geno:11}.
\STErev{In this scheme, the qubit is encoded in the polarization degree
of freedom of a light beam, whose intensity (photon) degree of freedom
has been prepared in a coherent state,} and the one-parameter gate
corresponds to a polarization transformation.  In the ideal case,
orthogonal polarization states can be perfectly discriminated.  However,
in a realistic scenario and especially in free-space communication,
non-dissipative (diffusive) noise affecting light polarization disturbs the
orthogonality of the states at the receiver, thus requiring suitable
detection and strategy for discrimination. It is worth noting that
coherent states preserve their fundamental properties when propagating
in purely lossy channels, suffering only attenuation, thus only the
noise affecting the polarization is detrimental. Remarkably, since our
receiver is phase-insensitive, the scheme works as well as when
phase-diffusion noise is affecting the channel.  This also holds in the
case of phase-randomized coherent states \cite{lab13} which can be
easily generated, characterized and manipulated \cite{lab14} and are
useful for enhancing security in decoy state quantum key distribution
\cite{lab15,lab16}.
\par
The paper is organized as follows; the physical system is described in
Section \ref{sec:2}, where the corresponding channel model and
log-likelihood metric are also defined. The associated channel capacity 
is evaluated in Section \ref{sec:3}, while the achievable residual frame and bit error rate  
obtained with LDPC coding is presented in Section \ref{sec:4}. Section 
\ref{sec:5} concludes the paper with some final remarks.
%%%
\section{The physical channel}\label{sec:2}
\begin{figure}[t!]
\vspace{-1.5cm}
\includegraphics[width=.98\columnwidth]{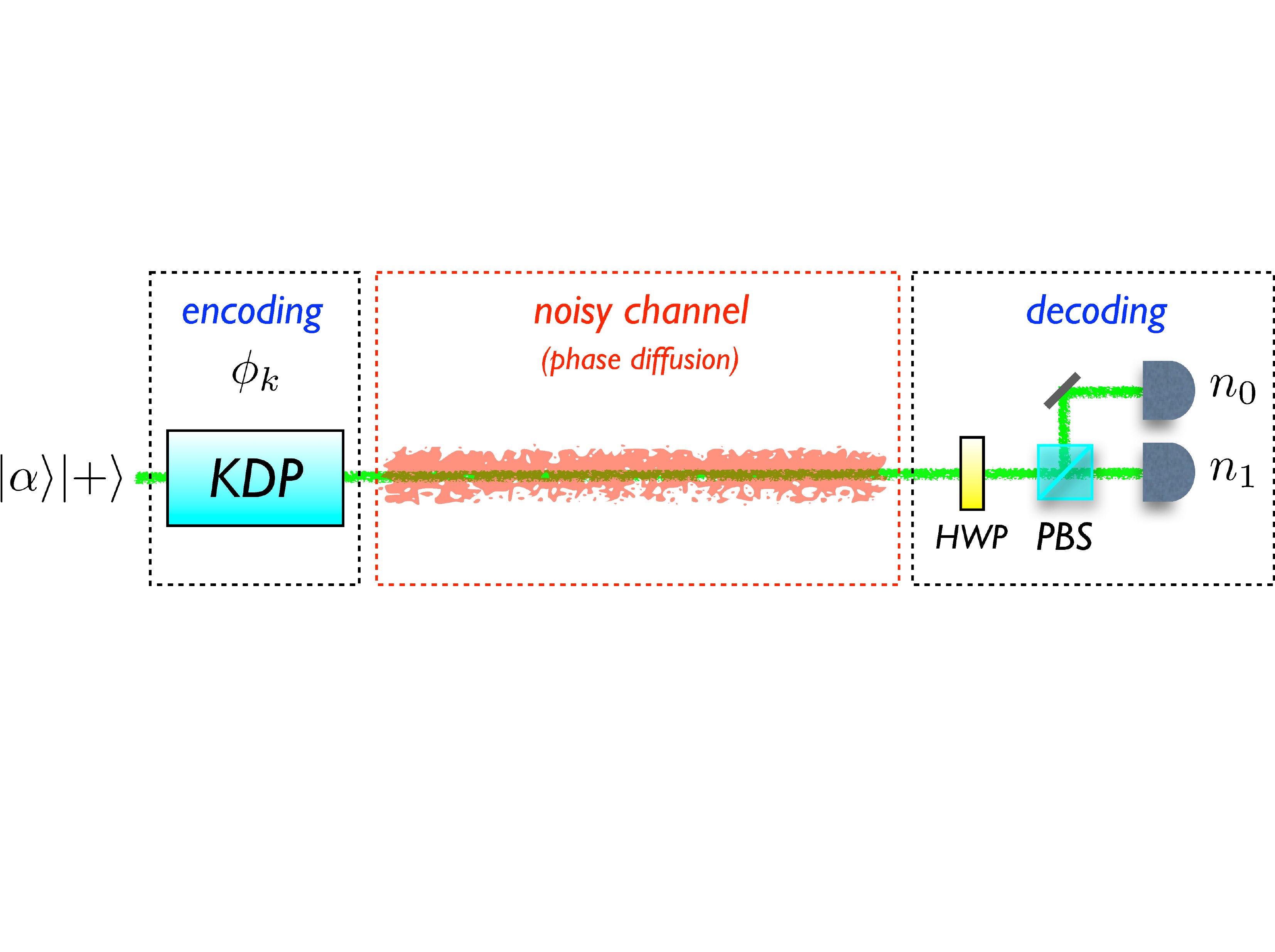}
\vspace{-2.2cm}
\caption{Schematic diagram of the optical setup implementing
photon-number-assisted, polarization-based, binary communication
channels equipped with photon counting receivers.}
\label{fig:1}
\end{figure}

The channel we are going to investigate corresponds to
the optical setup schematically depicted in Fig.~\ref{fig:1}.
\STErev{
The information bit  is encoded onto the polarization
degree of freedom of a  light beam prepared in a coherent state $\ket{\alpha}$,
initially linearly polarized at $45^\circ$ with respect to the $x$-axis, i.e.:
$$\ket{\alpha}\otimes\ket{+}=
\ket{\alpha}\otimes\left(\frac{\ket{H}+ \ket{V}}{\sqrt{2}}\right) 
\,,$$
where $|H\rangle$ and $|V\rangle$ denote horizontal and vertical 
polarization states with respect to the $x$-axis. The 
encoding rule for the bit $k=0,1$ is applied to the qubit by means of
the polarization rotation $U(\phi_k)=e^{-i\frac12 \phi_k \sigma_3}$,
$\sigma_3$ being the Pauli matrix. Due to the analogy with
the phase-shift encoding, from now on we will refer to $U(\phi_k)$
as ``phase shift''. In order to follow the  scheme proposed in Refs.~\cite{lab1}
and in view of a possible experimental verification reported in
\cite{lab12,geno:11}, we assume
that the encoding rule for the bit given in Table~\ref{t:encoding}.}
\begin{table}[h!] 
\begin{center}
{\large
\begin{tabular}{ccc}
\hline
$k$  & $\longrightarrow$  &  $\phi_k$\\
\hline
$0$   & $\longrightarrow$ &  $\pi/4$ \\
$1$   &  $\longrightarrow$ &  $3\pi/4$\\
\hline
\end{tabular}
}
\end{center}
\caption{Encoding rule for the polarization phase-shift. \label{t:encoding}}
\label{tab:1}
\end{table}
\par
\begin{figure}[h!]
        \centerline{\includegraphics[width=.7\columnwidth]{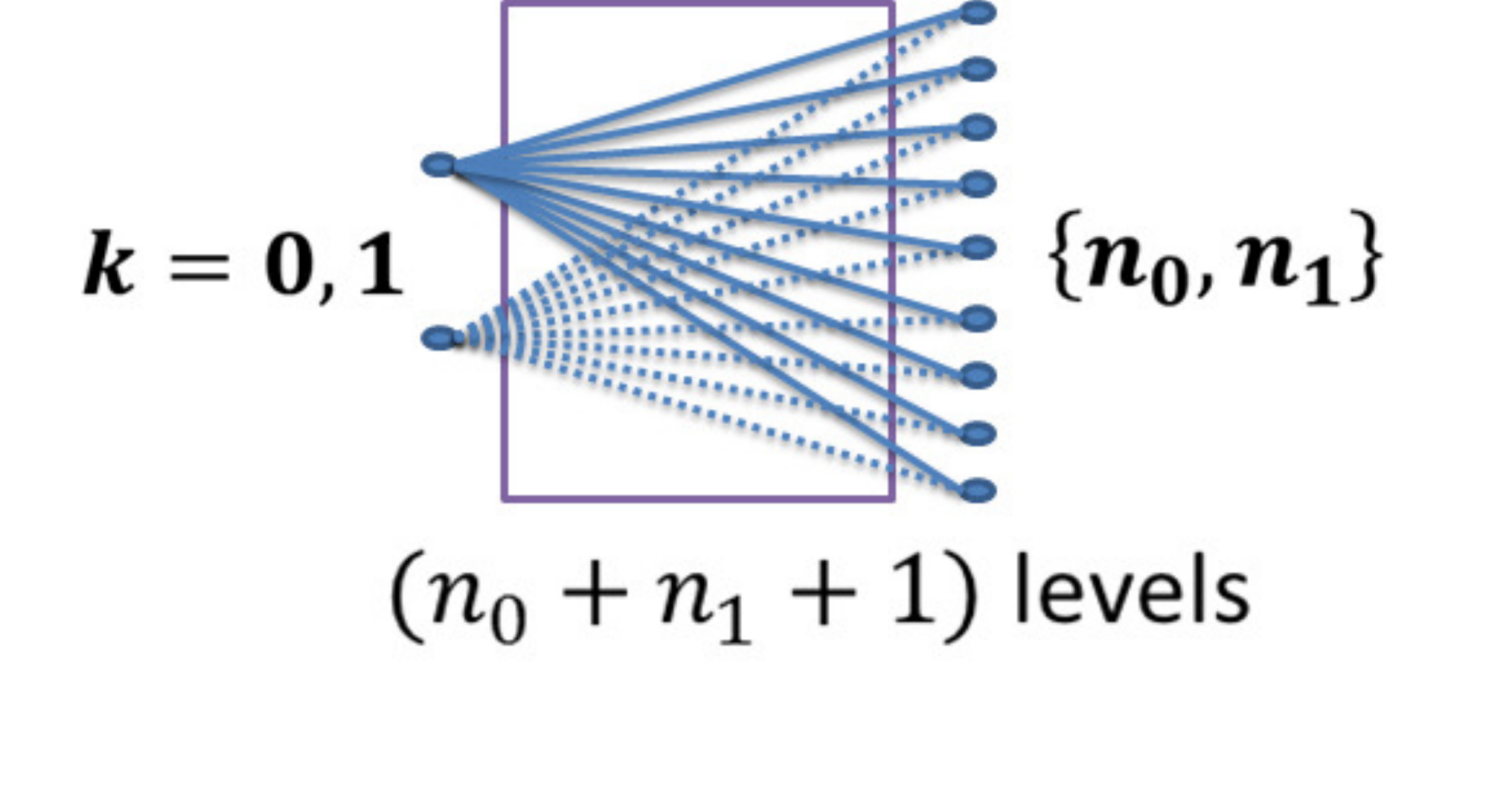}} 
    \vspace{-0.5cm}
    \caption{BIMO channel model of the considered system.}
    \label{fig:2}
\end{figure}

The polarization rotation (phase shift)
may be easily implemented by means of 
a KDP crystal driven by a high voltage generator, and corresponds to a change 
of the polarization from linear to elliptical.
At the detection stage information is retrieved by intensity
measurement, in a scheme involving a Half-Wave Plate (HWP), a Polarizing
Beam Splitter (PBS) and two photon counters. \STErev{This scheme has been 
experimentally tested to achieve one-parameter qubit gate optimal estimation
\cite{lab12,geno:11}. Furthermore, several examples of detectors now used
by the quantum optics community, can be used as photon counters
\cite{bondani:09,allevi:09,allevi:10,lab12,lab13,lab14}}.
The outcomes of the
measurement are thus pairs of integer numbers $(n_0,n_1)$, where $n_k$ is
the number  of detected photons in the reflected $(k=0)$ and transmitted $(k=1)$
beam, respectively. Notice
that the total number of detected photons $n=n_0+n_1$ is varying shot by shot. 
We assume that no photon is lost at the beam splitter. 
The number of photons in the coherent carrier is a Poisson distributed 
random variable with mean value $N_c=|\alpha|^2$. Also the two beams after
the PBS are coherent states and the joint probability of obtaining the
outcome $(n_0,n_1)$ is the product of two factorized Poisson distributions. 
The mean values depend on the polarization phase-shift, i.e. on
the bit value. Upon denoting by $N_k (\phi)$ the mean photon number
in the reflected or transmitted beam when the imposed 
phase-shift is $\phi$, we have: 
\begin{align}
N_0 (\phi) &= \frac12 N_c\, (1+\cos\phi),\quad
N_1 (\phi) &= \frac12 N_c\, (1-\cos\phi)\,. \notag
\end{align}
The probability of the event $(n_0, n_1)$ is thus given by:
\begin{align}
p(n_0,n_1|\phi)&= e^{-N_0(\phi)-N_1(\phi )} 
\frac{N_0(\phi)^{n_0}}{n_0!}
\frac{N_1(\phi)^{n_1}}{n_1!} \notag \\[1ex]
&= e^{-N_c} 
\frac{N_0(\phi)^{n_0}}{n_0!}
\frac{N_1(\phi)^{n_1}}{n_1!}\,.\label{probs}
\end{align}
The overall scheme is suitable for working with weak optical
signals, where the value of $N_c$ is typically small.
The relevant observation to be made here is that the information is retrieved 
by photon counting, and therefore the discrete bit value $k$, encoded 
in the polarization qubit, is mapped at the detection stage onto 
pairs of integer numbers.
The considered scheme can be modeled as shown in Fig.~\ref{fig:2}, i.e.
with an equivalent binary-input/multiple-output channel that receives
the binary random variable $k$ as input, and generates the two random
variables  ${n_0,n_1}$ as outputs. In particular, for a given number $n$
of detected photons, there are $n+1$ pairs ${n_0,n_1}$ such that
$n_0+n_1=n$.                                    
The availability of multiple outputs, whose likelihood can be exploited
for soft-information processing, is a crucial characteristic of the
described scheme.  
\par
\STErev{If propagation of the light beam occurs in an 
environment, which perturbs the polarization but preserves the energy,
then the state impinging onto the PBS has no longer a 
well-defined polarization (phase): If the initial state 
is $\ket{\phi_k}\otimes\ket{\alpha}$, 
where $\ket{\phi_k} = U(\phi_k) \ket{+}$ refers to 
the polarization qubit, the phase-diffusion noise affects 
the polarization according to the map \cite{lab12}:
\begin{align}\label{noise}
\ket{\phi_k} \to \varrho_k = 
\int_{\mathbbm R} d\varphi\, g(\varphi,\Delta)\,
U(\varphi) \ket{\phi_k} \bra{\phi_k} U^{\dag}(\varphi),
\end{align}
where $\varrho_k$ represents the density matrix of the degraded polarization qubit and $g(\varphi,\Delta)$ is a normal distribution of the variable $\varphi$ with zero mean and standard deviation $\Delta$. From the physical point of view, Eq.~(\ref{noise}) follows from a Master equation approach \cite{zoller} which represents a dynamics in which the quantum state of light undergoes an energy conserving ÒscatteringÓ affecting the polarization. Overall, this corresponds to applying a random polarization rotation (or phase shift) of the input polarization distributed according to $g(\varphi,\Delta)$.}
The probabilities of the outcomes are still given by Eq.~(\ref{probs}),
however with the mean photon numbers modified to:
\begin{align}\label{avD:1}
N_0 (\phi,\Delta) &\equiv
N_0 (\phi) =  \frac12 N_c\, (1+e^{-\Delta^2}\cos\phi),\\[1ex]
N_1 (\phi,\Delta) &\equiv
N_1 (\phi) =  \frac12 N_c\, (1-e^{-\Delta^2}\cos\phi)\,.
\label{avD:2}
\end{align}
%%%
\section{Evaluation of the Log-Likelihood Ratios}
\label{sec:2.1}
Soft-decoding algorithms are typically based on the use of Log-Likelihood-Ratios (LLR). 
In our particular case, the LLR values associated to the channel model of Fig.~\ref{fig:2} can be evaluated as:
\begin{equation}
\LLR(n_0,n_1)= \log_2\left[ 
\frac{p(\phi_1|n_0,n_1)}{p(\phi_0|n_0,n_1)}
\right]
\label{eq:1}
\end{equation} 
 where,
\begin{equation}
p\left(\phi_k|\{n_0,n_1\}\right) \quad  k=0,1
\label{eq:2}
\end{equation}
is the probability that the transmitted bit was ``$k$" given the 
outcomes $(n_0,n_1)$. Using Bayes theorem, Eq.~(\ref{eq:1}) may
be rewritten as:
\begin{align}
\LLR(n_0,n_1)= \log_2\left[ 
\frac{p(n_0,n_1|\phi_1)}{p(n_0,n_1|\phi_0)}
\right]
\label{eq:3}\,.
\end{align} 
Finally, using Eq.~(\ref{probs}) we arrive at:
\begin{equation}
\LLR(n_0,n_1 )= (n_0-n_1 )  \log_2 \left(\frac{q}{1-q}\right)
\label{eq:8}
\end{equation}                                                            
where,
\begin{align}
q& = 
\frac{1}{2}\left [1-e^{-\Delta^2 }  \cos\left(\frac{\pi}{4}\right)\right]	,
\end{align}
for the chosen encoding.
The system described up to this point represents, for a given $n$, a 
BIMO Discrete Memoryless Channel (DMC) \cite{thomas}
with binary input $k$ and $n+1=n_0 + n_1 + 1$ outputs $(n_0,n_1)$, 
where $n$ is a Poisson distributed random variable. In the next 
Section we will evaluate the capacity of this channel.
\section{Evaluation of capacity}
\label{sec:3}
A sufficient statistic for detection with photon
counting detectors is the difference photocurrent at the output, i.e. 
$D=n_1-n_0$. Since the two random variables $n_0$ and $n_1$ 
are Poisson distributed, the outcome $d$ of $D$ is Skellam distributed,
namely:
\begin{align}
p_D(d|\phi) = e^{-N_c} \left[\frac{N_1(\phi)}{N_0(\phi)}\right]^{d/2} 
I_{|d|} \left(2 \sqrt{N_1(\phi) N_0(\phi)}\right)\,,
\end{align}
where $I_{m}(z)$ is the modified Bessel function of the first kind,
such that:
\begin{align}
p_D(d|\phi_k) = e^{-N_c} \left(\frac{q}{1-q}\right)^{(-1)^k d/2} 
I_{|d|} \left(2 N_c \sqrt{q(1-q)}\right)\,.
\end{align}
Upon denoting by $\Phi$ the input binary variable, the relevant figure
of merit to evaluate the channel capacity is the mutual information: 
$$
I(\Phi,D) = H(\Phi) - H(\Phi|D)\,,
$$
where, 
$$H(\Phi)=-z_0 \log_2 z_0 - z_1 \log_2 z_1\,,
$$
is the Shannon entropy of the input alphabet, $z_0$ ($z_1 = 1 - z_0$)
being the a-priori probability of sending the bit $k=0$ ($k=1$) and
$H(\Phi|D)$ is the conditional entropy:
$$
H(\Phi|D) = - \sum_{k,d} p_D(d) p(\phi_k|d) \log_2 p(\phi_k|d)
$$
and, 
\begin{align}
p_D(d) & =z_0 p_D(d|\phi_0) + (1-z_0) p_D(d|\phi_1) 
\end{align}
is the overall probability of the outcome $d$, irrespective of the input
bit.
\par
Our BIMO DMC is neither symmetric nor weakly symmetric. Recall that a DMC is said to be symmetric if the rows
(and the columns) of the channel transition probability matrix are permutations of each other. 
If, on the other hand, the rows are permutations of each other and the
column sums are equal but the columns are not permutations of each
other, the DMC is said to be weakly symmetric. It can be shown that for
symmetric or weakly symmetric channels uniform probability on input
maximizes the mutual information thus yielding capacity.  However, it can be easily shown that the input probability distribution maximizing 
the mutual information in the BIMO case above is the uniform one, i.e.
$z_0=z_1=1/2$. The channel capacity is thus given by: 
\begin{align}
{\cal C} &= \max_{z_0} I(\Phi|D) \notag \\ 
&=1 +  \frac12 
\sum_{k,d} [p_D(d|\phi_0) + p_D(d|\phi_1)] p(\phi_k|d) \log_2 p(\phi_k|d)
\,.
\end{align}
Our goal is now to compare the capacity of the present
photon counting receiver channel to that of the
equivalent binary symmetric channel resulting from the detection
of optical signals by on/off receiver, which just discriminates the 
presence or the absence of radiation (i.e., performs hard decoding).
The transition probability of the equivalent BSC associated with the considered photon counting receiver (i.e. the raw BER, denoted in the following as QBER) can be obtained as:
\begin{align}
\QBER&=\sum\limits_{m=1}^{\infty}
{p_D(m|\phi_0 )}+\frac{1}{2} p_D(0|\phi_0 ), \\[1ex]
          &=\sum\limits_{m=1}^{\infty}{p_D(-m|\phi_1 )}+\frac{1}{2} p_D(0|\phi_1 ).            
\label{eq:9}
\end{align}
Essentially, assuming $\phi_0$ is true, a detection error occurs for a hard
decision detector if $D=n_1 -n_0 >0$. In case $D=0$, the detector can toss a fair coin and assign a decoded bit arbitrarily, 
in which case the probability of error is $\frac{1}{2} p_D(0|\phi_1 )$.  

%In order to assess the performance of a photon counting receiver, let us observe that the scheme that most resembles the proposed binary communication scheme is the Binary Phase Shift Keying (BPSK) scheme based on the use of coherent states. For such a scheme, nearly optimal performances are achieved by the Dolinar receiver with capacity \cite{lab9}:
%\begin{equation*}
%C_{\hbox{\footnotesize BPSK-Dolinar}}=1-H_2 \left[\frac12 \left(1-\sqrt{1-e^{-4N_c} }\right)\right]
%\end{equation*}
%where, $H_2$ is the binary entropy function (i.e., $H_2 (p)=-p\log_2 (p)-(1-p)\log_2 (1-p)$).
%This capacity is close to the ultimate capacity obtained using an as yet unknown optimal receiver:
%\begin{equation*}
%C_{\hbox{\footnotesize BPSK-Ultimate}}=1-H_2 \left[\frac12 \left(1+e^{-2N_c} \right)\right].
%\end{equation*}
%%
%We must however observe that the Dolinar receiver requires a complicated feedback system for its implementation, and it has therefore a significantly higher complexity with respect to a photon counting receiver 
%as presented here \cite{lab9}.

\begin{figure}[h!]
\centerline{\includegraphics[width=\columnwidth]{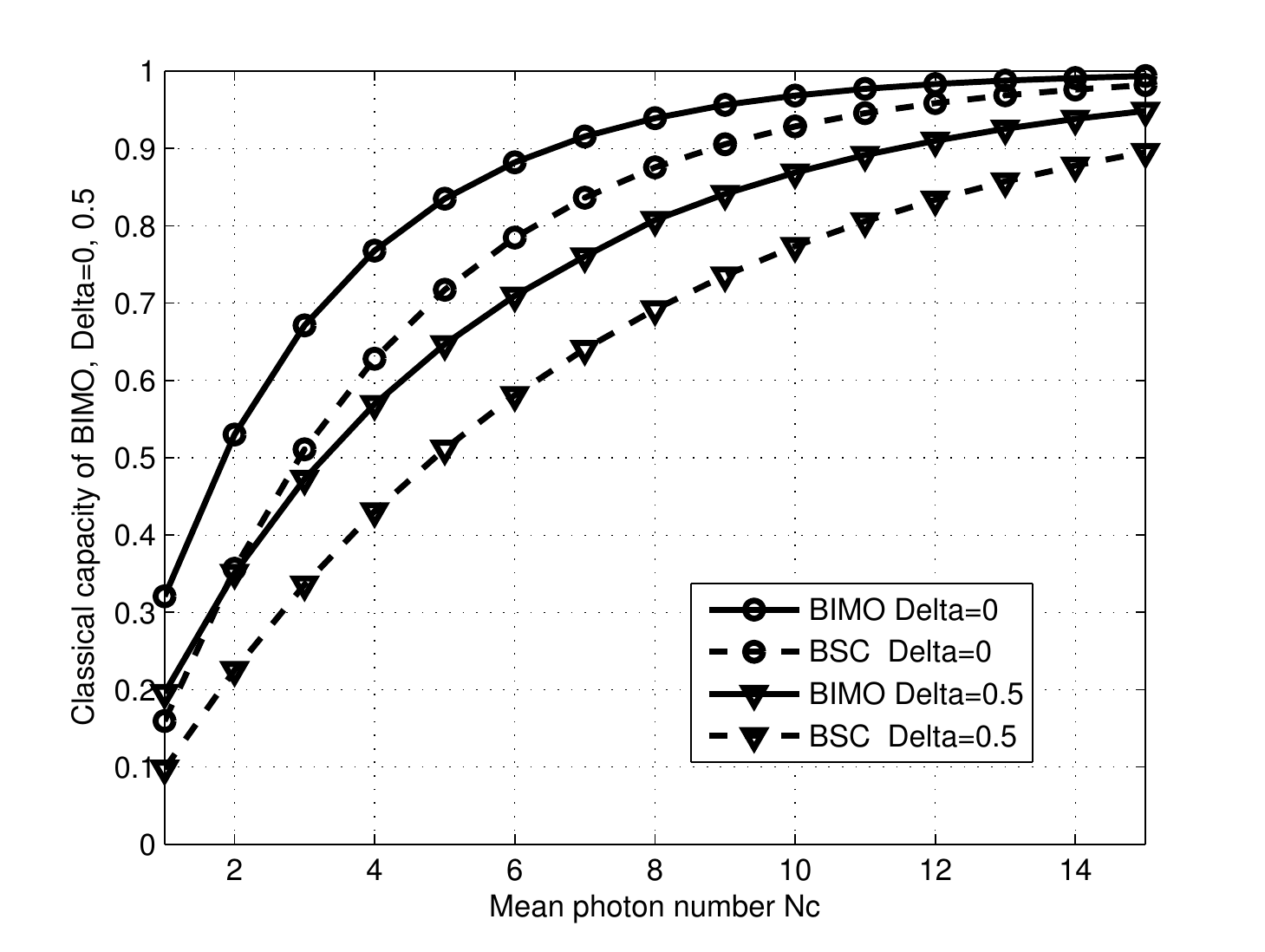}} 
\caption{Classical capacity  of the equivalent BSC with cross-over probability QBER (dashed curves) compared to that of the BIMO DMC (solid curves) for $\Delta=0$ (circle) and $\Delta=0.5$ (triangle) as a function of mean photon number $N_c$.}
\label{fig:3}
\end{figure}

\begin{figure}[h!]
\centerline{\includegraphics[width=\columnwidth]{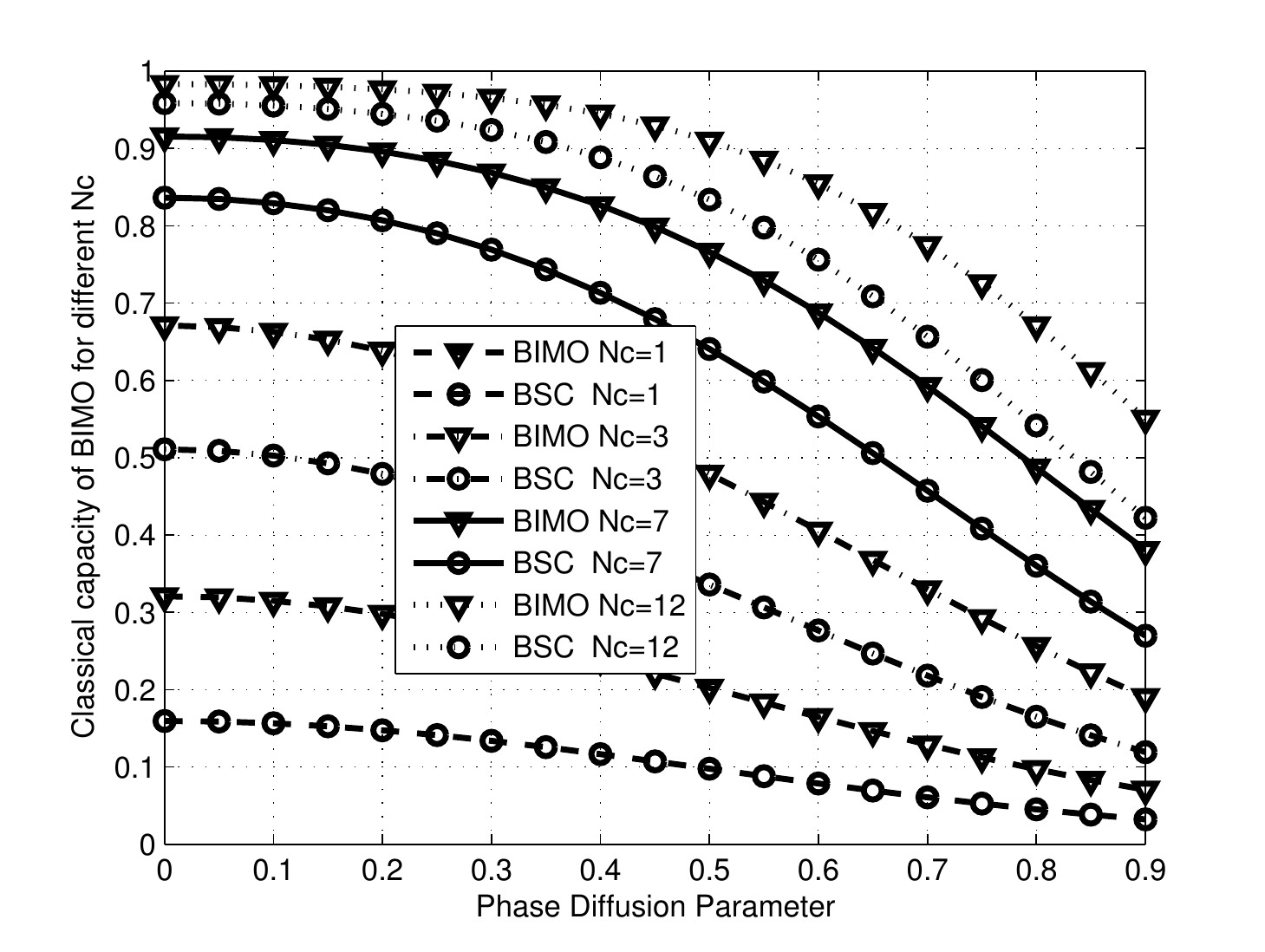}} 
\caption{Classical capacity of equivalent BSC with cross-over probability QBER (circle) and of BIMO DMC (triangle) for $N_c=1,3,7,12$ as a function of the phase diffusion parameter $\Delta$.}
\label{fig:4}
\end{figure}

In our case, in the limit $N_c \gg1$, we can write:
\begin{equation*}
\frac{N_1(\phi_k)}{N_2(\phi_k)  } = \frac{p(1|\phi_k)}{p(0|\phi_k)} 
\end{equation*}
and,
\begin{equation*}
N_1(\phi_k)\,  N_2(\phi_k)=N_c^{2} \, p(1|\phi_{k})\, p(0|\phi_{k}).
\end{equation*}
When ``$0$ is transmitted and it is mapped to $\phi_0$", we get from Eqs. (\ref{avD:1})
and (\ref{avD:2}): 
\begin{align*}
p_D(m &|\phi_0 )=\frac{e^{-N_c}}{ \sqrt{\alpha_{\Delta}^m}} \;
B_m(N_c,\Delta);
\end{align*}
analogously when ``1 is transmitted and it is mapped to  $\phi_1$" :
 \begin{align*}
p_D(m &|\phi_1 )= e^{-N_c} \sqrt{\alpha_{\Delta}^m}\;
B_m(N_c,\Delta),
\end{align*}
where,
\begin{equation*}
\alpha_{\Delta}=\frac{\sqrt{2}+e^{-\Delta^2}}{\sqrt{2}-e^{-\Delta^2}},
\end{equation*}
and,
\begin{align*}
B_m(N_c,\Delta)=I_{|m|}  \left(N_c\sqrt{1-\frac12\, e^{-2\Delta^2}}\right).
\end{align*}
%Let $X$ be the random variable associated with the transmitted phase $\phi_k$ (or the transmitted logical bit ``$k$") and $Y$ be the channel output, 
%i.e. our sufficient statistic $n_1-n_0$. Then, the formulas above give us the channel transition probabilities for the
%considered DMC. Using the classic definition of mutual information:
%\begin{equation*}
%I(X;Y)=H(X)-H(X|Y)
%\end{equation*}
%and noting that the input is binary with $p(X=0)=p(\phi_0 )=\p$, and $p(X=1)=p(\phi_1 )=1-\p$, 
%
After some manipulation we have:
\begin{align*}
p_{0,m}\equiv p(\phi_0 |D=m)=\frac{z_0}{z_0\, (1-\alpha_{\Delta}^m )+\alpha_{\Delta}^m},\\[1ex]
p_{1,m}\equiv p(\phi_1 |D=m)=\frac{1-z_0}{z_0\, (1-\alpha_{\Delta}^m )+\alpha_{\Delta}^m}.
\end{align*}
The final expression of the conditional entropy as a function of the two parameters $N_c$ and $\Delta$ is:\\
\begin{align*}
H(\Phi|D)&=-e^{-N_c}\sum_m{\frac{z_0\, B_m(N_c,\Delta)}{\sqrt{\alpha_{\Delta}^{m}}}\;
\log_2 \left(p_{0,m}\right)}\\
&-e^{-N_c}\sum_m{(1-z_0)\, B_m(N_c,\Delta) \sqrt{\alpha_{\Delta}^{m}} \;
\log_2\left(p_{1,m}\right)}.
\end{align*}
Note that capacity is achieved with $z_0=\frac{1}{2}$. The results are
shown in Fig.~\ref{fig:3} and \ref{fig:4}. The capacity of the BIMO DMC compared to that of the equivalent BSC obtained in case of on/off detection is shown in Fig.~\ref{fig:3} as a function of the mean photon number $N_c$ and for $\Delta=0, 0.5$, while in Fig.~\ref{fig:4} the BIMO DMC is compared to the equivalent BSC for $N_c=1,3,7,12$ as a function of the phase diffusion parameter
$\Delta$. It is possible to observe that a higher capacity can be obtained by the BIMO DMC with respect to the equivalent BSC, that may possibly lead to improved BER improvement when an error correction code is applied to the two channels. This aspect is indeed investigated in the next section. 
The capacity improvement offered by the photon counting detector decreases as $N_c$ increases, in particular for low values of $\Delta$, as it can be observed by both Fig.~\ref{fig:3} and Fig.~\ref{fig:4}.

\section{BER PERFORMANCE IN PRESENCE OF FEC}
\label{sec:4}
This section investigates the performance obtainable with Forward Error Correction (FEC) codes applied to the scheme of Figs.~\ref{fig:1} and \ref{fig:2}. The $m$-bits codeword of a systematic FEC code with code rate $R_c$ is generated concatenating $L$ information bits and $r$ redundancy bits so that $m= L + r$ and $R_c=L/(L+r)$. 

\begin{figure}[h!]
        \centering
                \includegraphics[width=\columnwidth]{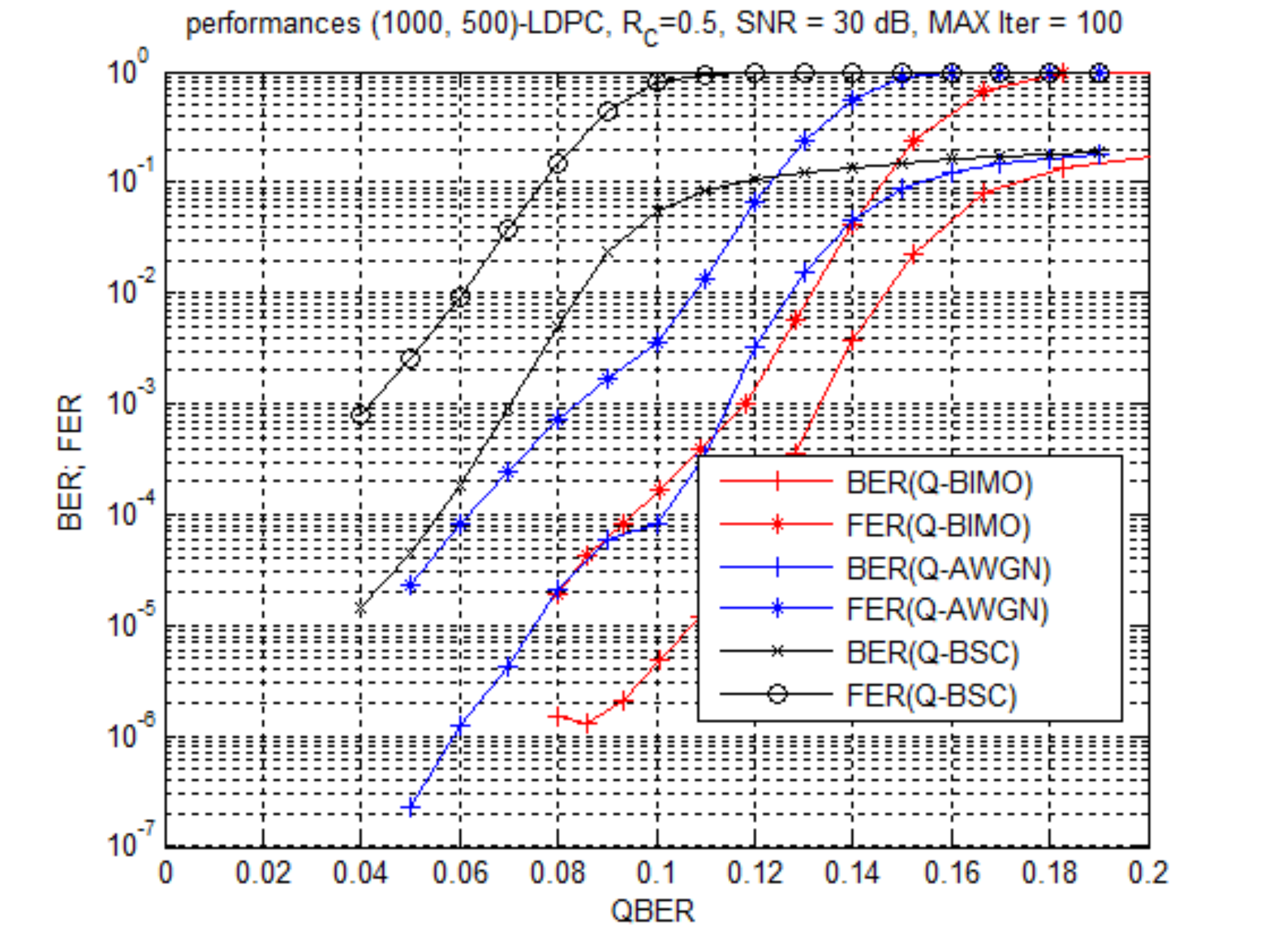}
                \includegraphics[width=\columnwidth]{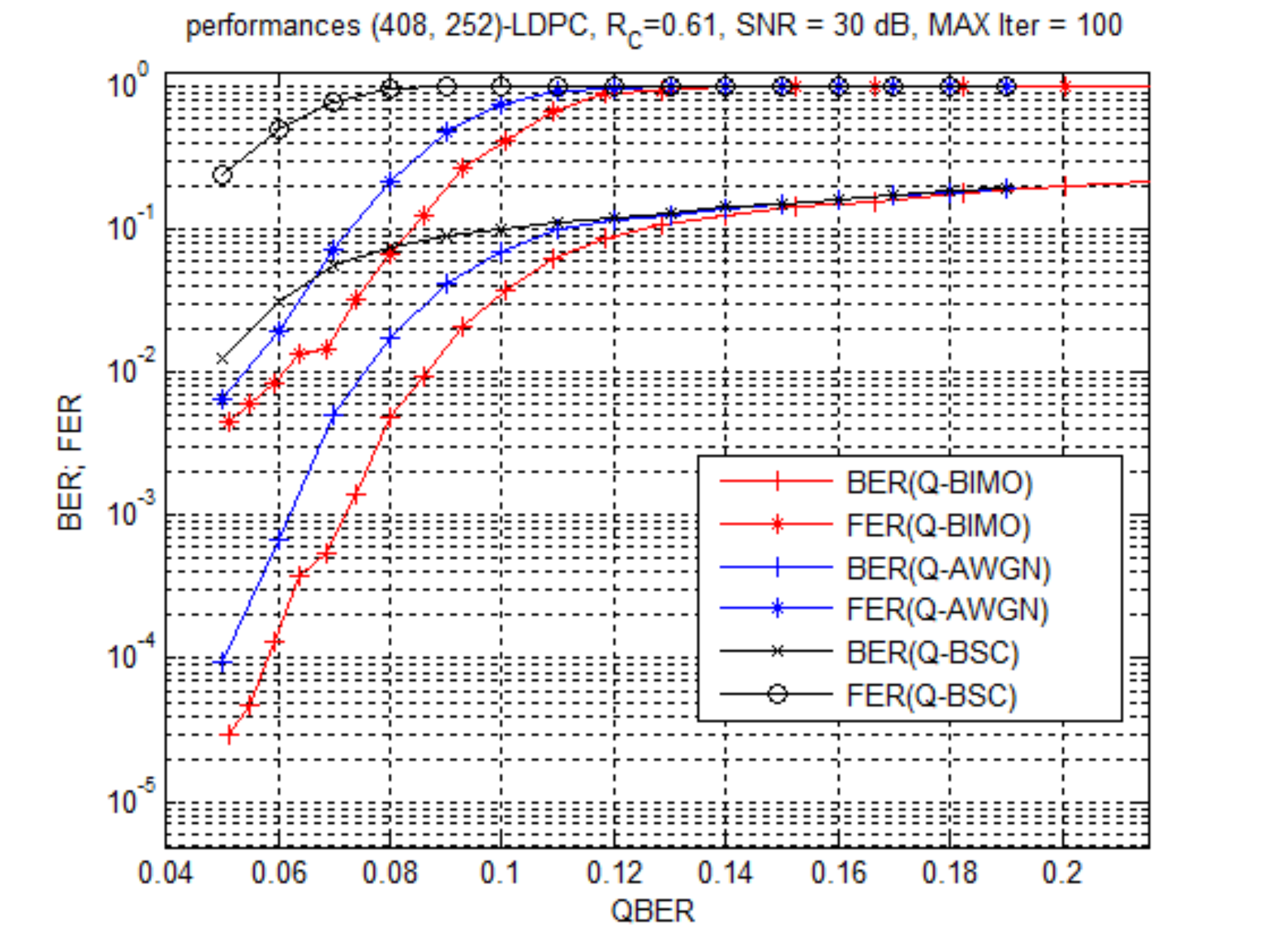}
                \includegraphics[width=\columnwidth]{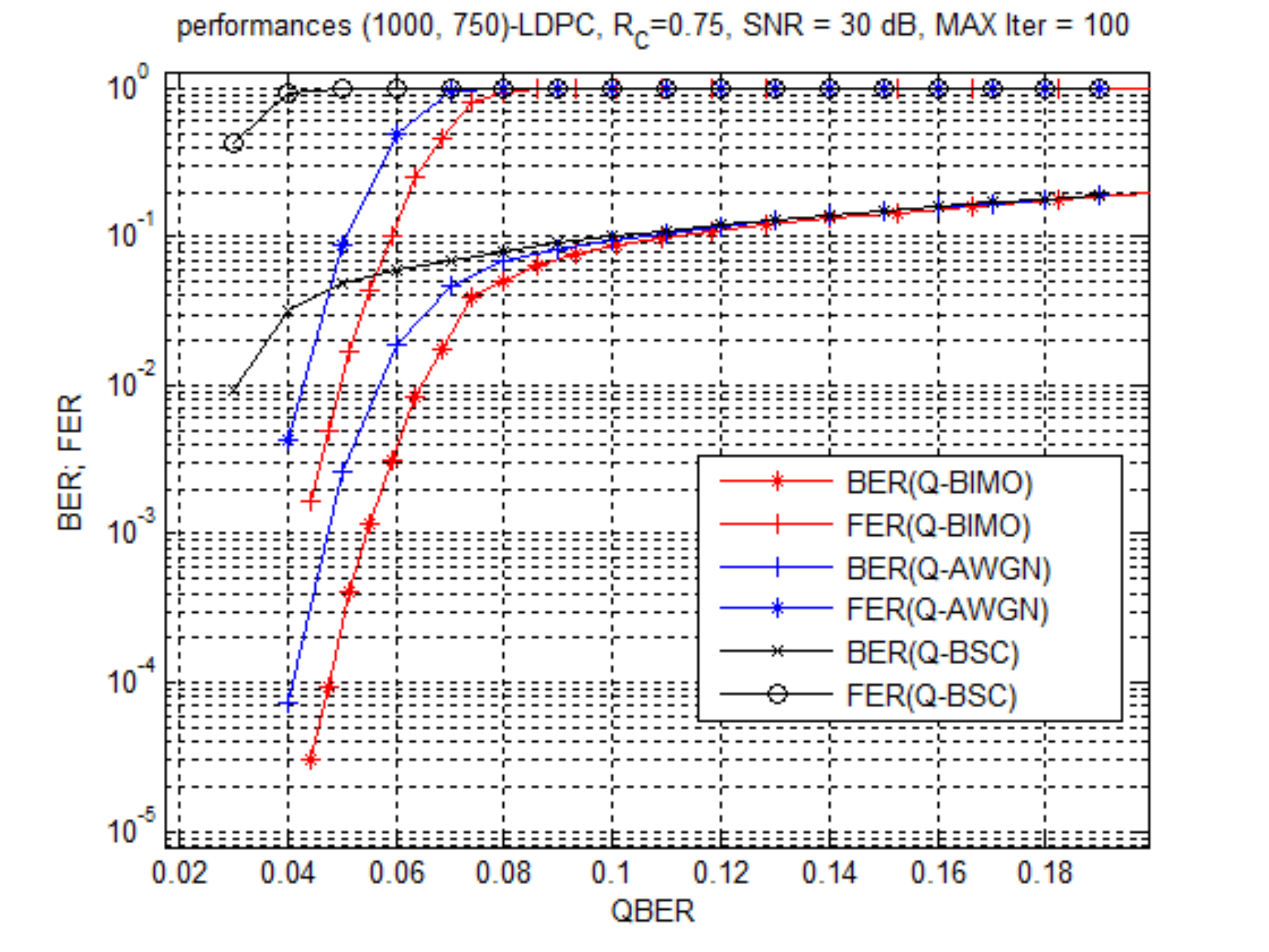}
        \caption{Simulated BER and FER values for a LDPC code with  $L=500, r = 500$ and $R_c = 0.5$ (top plot), $L=252$, $r = 156$ and $R_c = 0.6$ (center plot) and $L=750$, $r = 250$ and $R_c = 0.75$ (bottom plot), obtained with different models of the quantum channel: BSC (Q-BSC curves, black), AWGN (Q-AWGN curves, blue) and BIMO DMC (Q-BIMO curves, red).}
\label{fig:5}
\end{figure}

A systematic low density parity check code has been selected as test FEC code, due to its capacity achieving  performance (albeit at very large block lengths) and low complexity iterative decoding structure, and a simulation analysis has been performed to assess the potential performance improvements obtainable using the soft-metric of Eq.~(\ref{eq:8}). Three different quantum channel models have been considered, all with the same equivalent uncoded raw BER value, that will be denoted as QBER. The simulation results are shown in Fig.~\ref{fig:5}, where each pair of BER-FER curves depicts the  residual bit error rate and frame error rate after channel decoding. The following parameters have been considered:
\begin{itemize}
\item $R_c = 0.5 $, $L=500$, $r = 500$, 
\item $R_c = 0.61$, $L=252$, $r = 156$, 
\item $R_c = 0.75$, $L=750$, $r = 250$.
\end{itemize}

The black curves in Fig.~\ref{fig:5} labeled "Q-BSC" are associated with an equivalent BSC with binary input $X=k$, binary output $Y$ and transition probability QBER derived from Eq.~(\ref{eq:9})  (i.e. a receiver that does not use the additional information derived from the knowledge of $n_0$ and $n_1$ and simply performs on/off detection) with LLR values \cite{lab17}:
\begin{align*}
\LLR(Y)&=\log_2\left[\frac{P(Y=1|X)}{P(Y=0|X)}\right]\\[1ex]
&=
\begin{cases}
    \log_2\left(\frac{1-{\mbox{\footnotesize $\QBER$}}}{{\mbox{\footnotesize $\QBER$}}}\right) ,& \text{if } X = 1;\\[1ex]
    \log_2\left(\frac{{\mbox{\footnotesize $\QBER$}}}{1-{\mbox{\footnotesize $\QBER$}}}\right) ,& \text{if } X = 0.
\end{cases}
\end{align*}
The blue curves labeled as ``Q-AWGN" represent the performance obtainable over a fictitious Additive White Gaussian Noise (AWGN) channel model with a Signal to Noise Ratio (SNR) selected in order to achieve an uncoded bit error probability QBER with a binary antipodal scheme. The curves labeled as ``Q-BIMO" represent the main result and are obtained transmitting through the BIMO DMC quantum channel model shown in Fig.~\ref{fig:2} with equivalent uncoded bit error probability QBER and using as input soft-metrics for the LDPC decoder, the LLR values generated via photon counting according to Eq.~(\ref{eq:8}). 
% Notice that for the ``Q-BIMO" curves the QBER parameter is actually the cross-over probability of
% the equivalent BSC as defined in Eq. (\ref{eq:9}) for the BIMO channel.

As it is apparent from the results for the photon counting receiver,
the BER and FER performance largely improve when the BIMO DMC and the LLR metrics from
Eq.~(\ref{eq:8}) are employed instead of the simpler BSC metrics. As an example, in the upper plot in Fig.~\ref{fig:5} for $\QBER=0.1$ the BIMO DMC with soft-metric processing offers almost three orders of magnitude improvement in BER with respect to the BSC model and the associated hard-metric processing. We must note that the curves labeled as ``Q-AWGN" must only be used as reference, since with the small number of photons we considered in our simulations the AWGN channel model would not be
appropriate. 

A comparison among the residual FER and BER values obtainable with the considered channel models for LDPC codes with code rates $0.61$ (center plot) and $0.75$ (bottom plot) is shown in Fig.~\ref{fig:5}. Also in these cases, both FER and BER values improve up to several orders of magnitude when using a photon counting receiver and the associated LLR values. Furthermore, we can observe that as the code rate increases, the ``Q-BIMO" performances obtained with BIMO LLR metrics get closer to the ``Q-AWGN" performances obtained with classic AWGN LLR metrics (although, as mentioned before, the AWGN model is not applicable in case of low number of received photons).
	
\begin{figure}[h!]
\centerline{\includegraphics[width=\columnwidth]{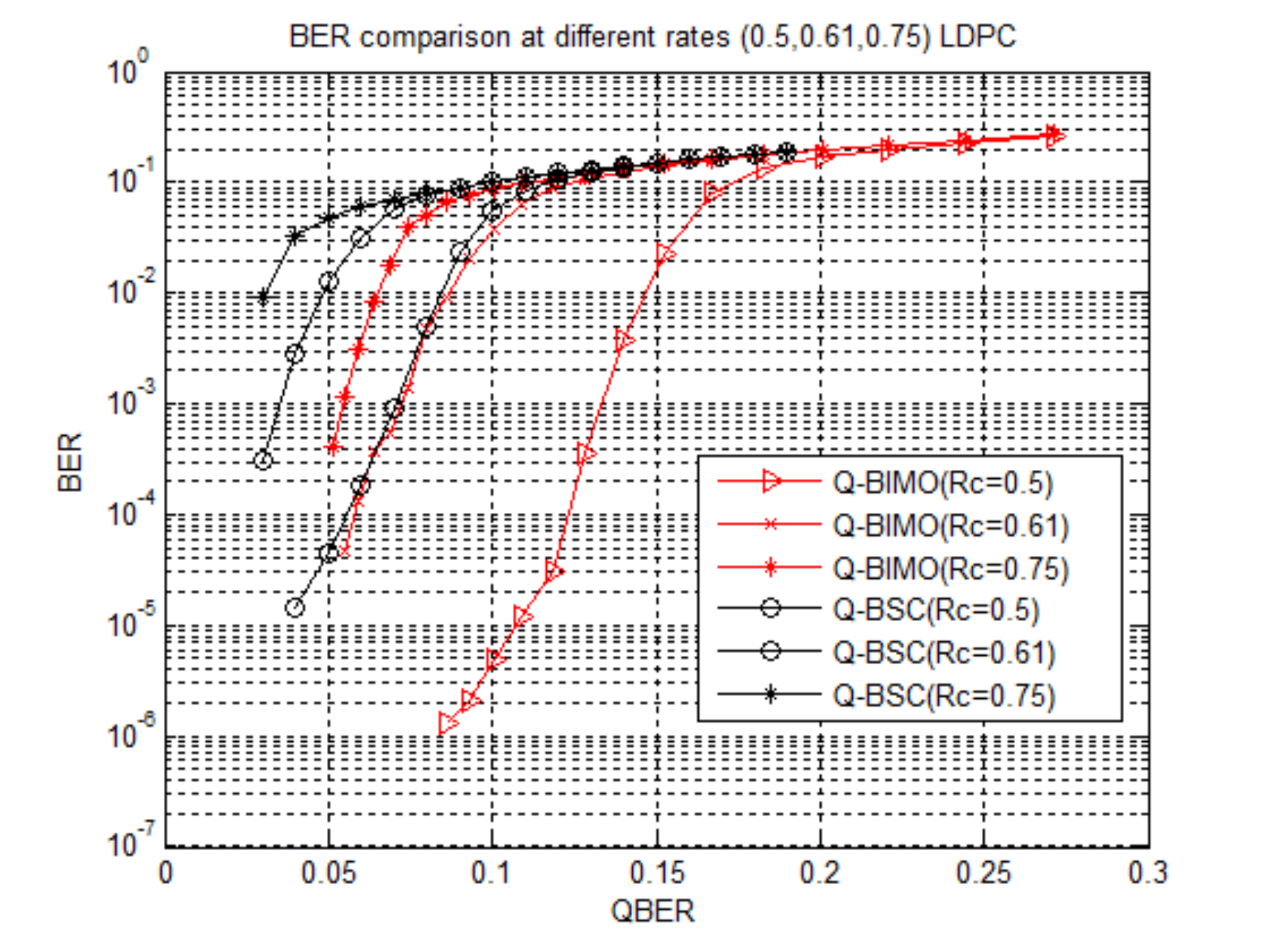}} 
\caption{Simulated residual BER obtained with BSC (Q-BSC curves, black) and BIMO DMC (Q-BIMO curves, red) models of the quantum channel and LDPC codes with different code rates ($R_c = 0.5,0.61$ and $0.75$).} 
\label{fig:6}
\end{figure}

Fig.~\ref{fig:6} compares the BER values obtained with the BSC and the BIMO channel models for different code rates, showing that, as expected, for higher rates, a lower QBER value is required before significant coding gains can be observed. 
 \begin{figure}[h!]
\centerline{\includegraphics[width=\columnwidth]{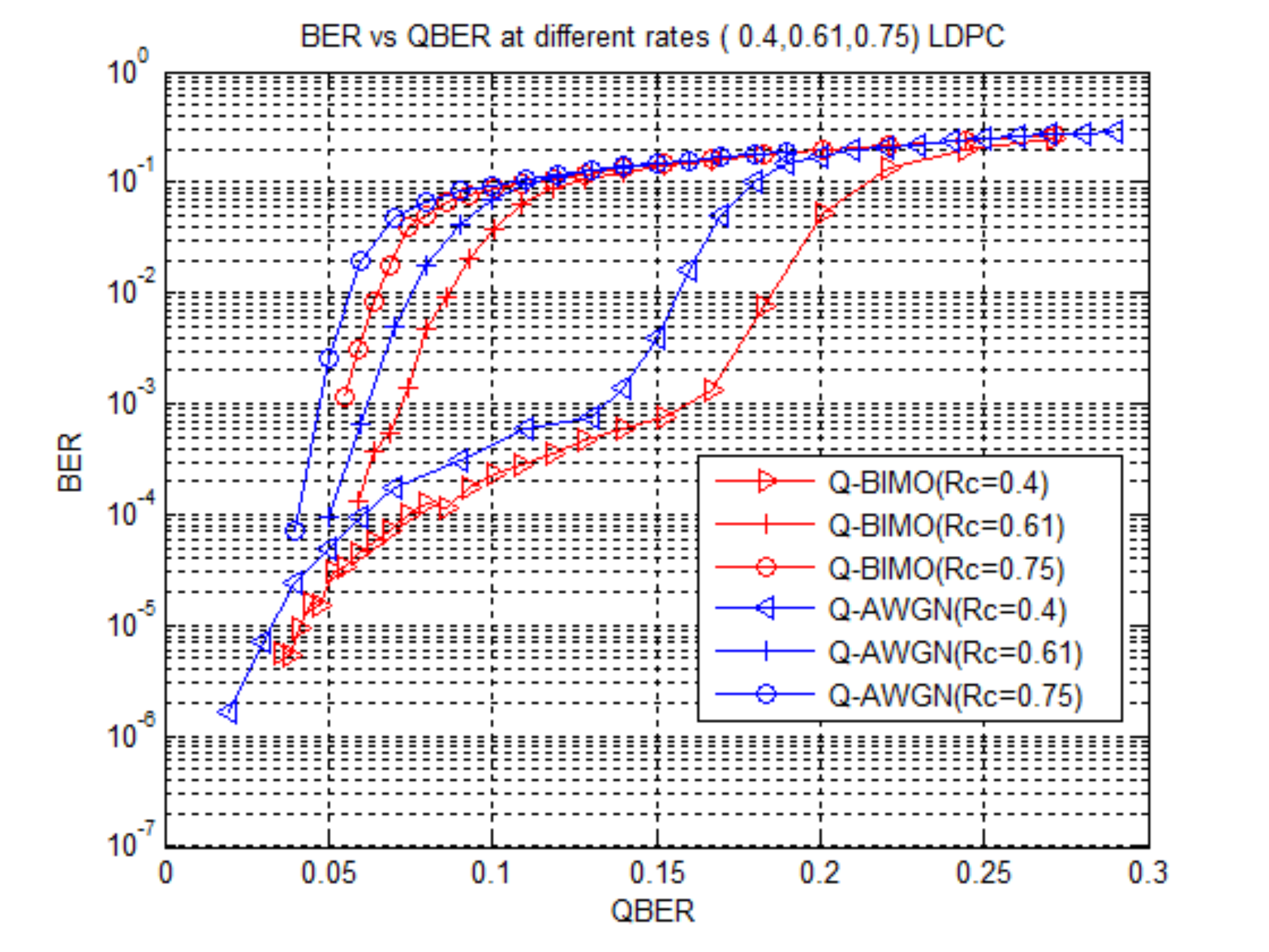}} 
\caption{Simulated residual BER obtained with AWGN (Q-AWGN curves, blue) and BIMO DMC (Q-BIMO curves, black) models of the quantum channel and LDPC codes with different code rates ($R_c = 0.4,0.61$ and $0.75$).}
\label{fig:7}
\end{figure}
>From Fig.~\ref{fig:7}, we can observe that for high values of $N_c$ (i.e. at low values of QBER) the BIMO DMC model can be approximated with an AWGN model, while the AWGN model approximation may be  unreliable at high QBER (low $N_c$) values, in particular at lower code rate values.
\begin{figure}[h!]
\centerline{\includegraphics[width=\columnwidth]{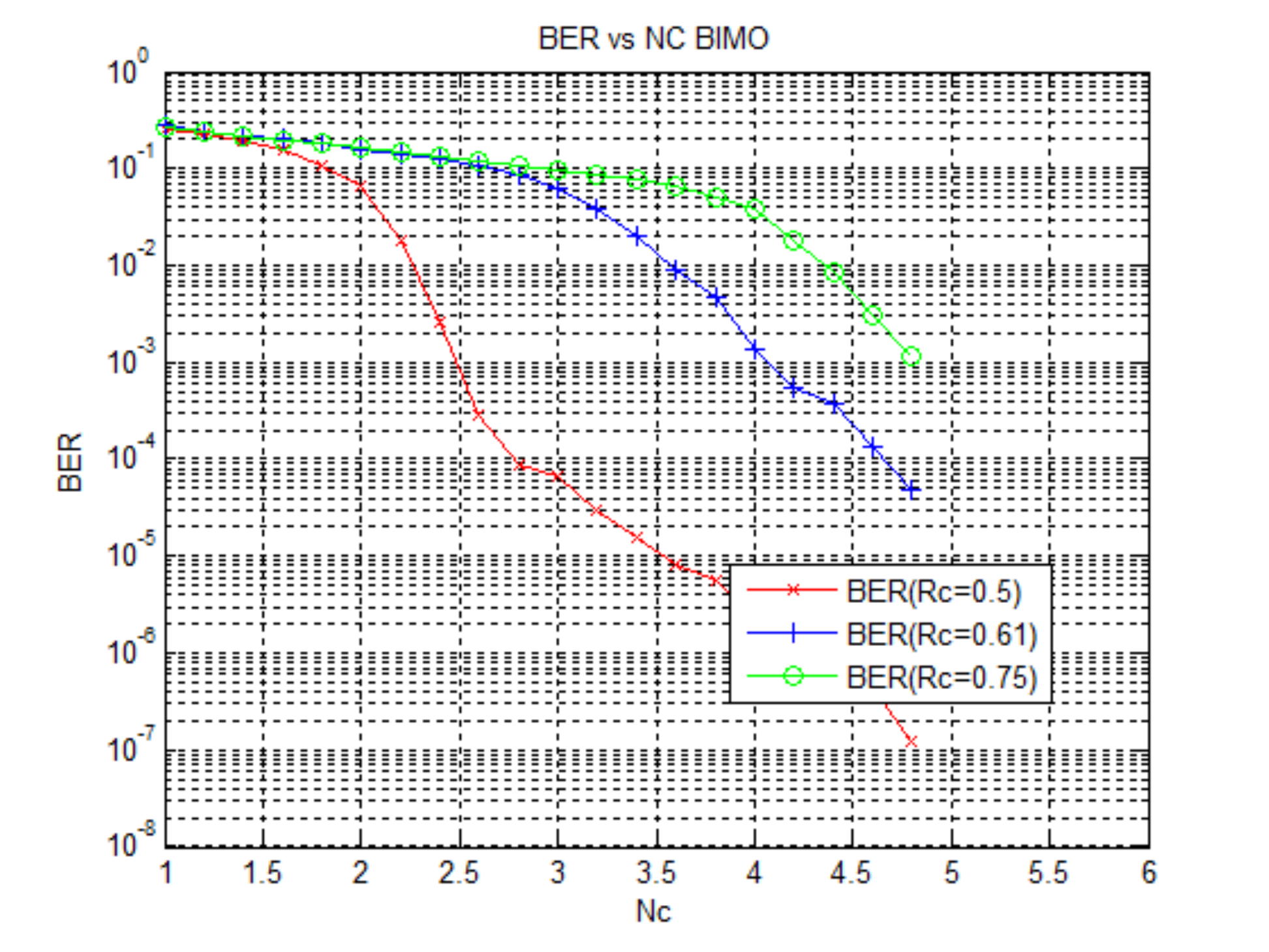}} 
\caption{Simulated residual BER for LDPC codes with $R_c = 0.5,0.61,0.75$ over BIMO DMC as a function of the mean photon number $N_c$.}
\label{fig:8}
\end{figure}
Finally, Fig.~\ref{fig:8} shows the residual BER obtained on the BIMO channel by LDPC codes with code rate $R_c = 0.5,0.61,0.75$ for different values of $N_c$.

\section{Conclusions}
\label{sec:5}
In this paper a photon-number-assisted, polarization-based binary transmission scheme equipped with a low-complexity photon counting receiver has been considered, analyzing both its capacity and its BER performance in presence of  
capacity achieving low density parity check codes. Different channel models 
applicable to the considered transmission scheme have been compared, 
proposing a time varying binary-input/multiple-output model and evaluating its LLR
metrics and channel capacity. It has been shown how the BIMO
channel model outperforms the corresponding BSC model, by taking
full advantage of the additional information offered by the photon counting detector.  
It was also shown that, as expected, the advantage offered by the photon counting detector deceases as the mean photon number $N_c$ increases, and that the BIMO model can be
approximated by an AWGN model at low values of QBER, i.e. for high values of $N_c$.
\section*{Acknowledgment}
This work was supported by MIUR (grant FIRB ``LiCHIS" - RBFR10YQ3H) and NATO (SfP project 984397 "Secure Quantum Communications").

%%%%%%%%%%%%%%%%%%%%%%%%%%%%%

\begin{IEEEbiographynophoto}{Marina Mondin} is Associate Professor at
Dipartimento di Elettronica, Politecnico di Torino. Her current
interests are in the area of signal processing for communications,
modulation and coding, simulation of communication systems, and quantum
communication. She holds two patents. She has been Associate Editor for
IEEE Transactions on Circuits and Systems-I from 2010 to 2013 and has
been in the 2012 TCAS committee for the selection of the Darlington and
Guillemin-Cauer Best Paper Awards. She is currently coordinating the
NATO SfP project ``Secure Communication Using Quantum Information
Systems". She is author of more than 150 publications. 
\end{IEEEbiographynophoto}

\begin{IEEEbiographynophoto}{Fred Daneshgaran} received the Ph.D. degree
in electrical engineering from University of California, Los Angeles
(UCLA), and since 1997 has been a full Professor with the ECE Department
at California State University, Los Angeles (CSLA). From 2006 he serves
as the chairman of the ECE department. Dr. Daneshgaran is the founder of
Euroconcepts, S.r.l, a R\&D company specializing in the design of
advanced communication links and software radio that operated from 2000
to 2010. From 1999 to 2001 he was the Chief Scientist and member of the
management team, for TechnoConcepts, Inc. where he directed the
development of a prototype software defined radio system, managed the
hardware and software teams and orchestrated the entire development
process. He is the director of the fiber and non-linear optics research
laboratory at CSLA, and served as the Associate Editor of the IEEE
Trans. On Wireless Comm.  in the areas of modulation and coding,
multirate and multicarrier communications, broadband wireless
communications, and software radio, from 2003 to 2009.  He has served as
a member of the Technical Program Committee (TPC) on numerous
conferences. Most recent contributions include IEEE WCNC 2014, CONWIRE
2012, ISCC 2011 to 2014, and PIMRC 2011.  
\end{IEEEbiographynophoto}

\begin{IEEEbiographynophoto}{Inam Bari} Inam Bari obtained his BS in
Telecommunication Engineering from the National University of Computer
and Emerging Science (NUCES-FAST), Pakistan in 2007, and was awarded
bronze medal. In 2008, he was awarded a full 5 years MS leading to PhD
scholarship by the Higher Education Commission of Pakistan. He obtained
his MS and PhD degrees from Politecnico di Torino, Italy, and is
currently  Assistant Professor at NUCES-FAST, Peshawar, Pakistan.
\end{IEEEbiographynophoto}

\begin{IEEEbiographynophoto}{Maria Teresa Delgado} received her BS
Degree in Electrical Engineering at the Universidad Central de Venezuela
in Caracas, Venezuela in 2006, and her MS and Ph.D. degrees in
Telecommunication Engineering from Politecnico di Torino, Italy, in 2008
and 2012. Her interests are in the area of signal processing for
telecommunications, coding, simulation of communication systems, quantum
cryptography and physical layer security for wireless and quantum
communication systems.  She is currently researcher at Istituto
Superiore Mario Boella, Turin, Italy.  
\end{IEEEbiographynophoto}

\begin{IEEEbiographynophoto}{Stefano Olivares} 
received the Ph.D. degree in Physics from the
University of Milan, Milano, Italy, and is currently a Researcher at
the Department of Physics, University of Milan, Italy. He is a
theoretician and his interests include quantum information, quantum
estimation, quantum optics, quantum interferometry and quantum
computation. 
His main
contributions are in the fields of quantum estimation of states and
operations, generation and application of entanglement, quantum
information, communication and decoherence.  Although his research
activity is mainly theoretical, he is an active collaborator in many
experimental groups. He is author of about 100 publications.
\end{IEEEbiographynophoto}

\begin{IEEEbiographynophoto}{Matteo G.~A.~Paris}
received his Ph.D. in physics from University of Pavia, and
is currently professor of quantum information and quantum optics 
at the Department of Physics of the University of Milan. His main
contributions are in the fields of quantum estimation of states and
operations, quantum tomography, generation, characterization and
application of entanglement, quantum interferometry, nonclassical states
and open quantum systems. In these fields he is author of about 250
publications in international journals. From 2013 he is editor-in-chief
of {\em Quantum measurements and quantum metrology}.
\end{IEEEbiographynophoto}

\vfill

\end{document}